\documentclass[prl, twocolumn, final, reprint, amsmath, amssymb, superscriptaddress, showpacs, a4paper, aps, 10pt, hidelinks, numbers, sort&compress, floats, balancelastpage]{revtex4-1}

\usepackage{graphicx, color} 
\usepackage[normalem]{ulem}          

\usepackage[colorlinks=true]{hyperref}
\hypersetup{pdfpagemode=None, linkcolor=black, citecolor=blue, urlcolor=blue}  

\usepackage[utf8]{inputenc}

\usepackage[normalem]{ulem} 
\usepackage{xcolor}

\newcommand{\1}{\ensuremath{\left|1 \right\rangle}}

\definecolor{britishracinggreen}{rgb}{0.0, 0.26, 0.15}
\definecolor{bulgarianrose}{rgb}{0.28, 0.02, 0.03}
\definecolor{darkred}{rgb}{0.90,0,0}
\definecolor{darkgreen}{rgb}{0,0.60,.2}
\definecolor{darkblue}{rgb}{0,0,1}
\definecolor{orange}{cmyk}{0,0.6,0.8,0}
\definecolor{lightblue}{rgb}{0.3,0.5,1}
\definecolor{lightgreen}{rgb}{0.4,0.80,.4}

\begin{document}

\title{Exploring ultracold collisions in $^6$Li-$^{53}$Cr Fermi mixtures: Feshbach resonances and scattering properties of a novel alkali-transition metal system}


\author{A. Ciamei}
\email{ciamei@lens.unifi.it}
\affiliation{Istituto Nazionale di Ottica del Consiglio Nazionale delle Ricerche (CNR-INO), 50019 Sesto Fiorentino, Italy}
\affiliation{\mbox{European Laboratory for Non-Linear Spectroscopy (LENS), Universit\`{a} di Firenze, 50019 Sesto Fiorentino, Italy}}
\author{S. Finelli}
\affiliation{Istituto Nazionale di Ottica del Consiglio Nazionale delle Ricerche (CNR-INO), 50019 Sesto Fiorentino, Italy}
\affiliation{\mbox{European Laboratory for Non-Linear Spectroscopy (LENS), Universit\`{a} di Firenze, 50019 Sesto Fiorentino, Italy}}
\author{A. Trenkwalder}
\affiliation{Istituto Nazionale di Ottica del Consiglio Nazionale delle Ricerche (CNR-INO), 50019 Sesto Fiorentino, Italy}
\affiliation{\mbox{European Laboratory for Non-Linear Spectroscopy (LENS), Universit\`{a} di Firenze, 50019 Sesto Fiorentino, Italy}}
\author{M. Inguscio}
\affiliation{Istituto Nazionale di Ottica del Consiglio Nazionale delle Ricerche (CNR-INO), 50019 Sesto Fiorentino, Italy}
\affiliation{\mbox{European Laboratory for Non-Linear Spectroscopy (LENS), Universit\`{a} di Firenze, 50019 Sesto Fiorentino, Italy}}
\affiliation{Department of Engineering, Campus Bio-Medico University of Rome, 00128 Rome, Italy}
\author{A. Simoni}
\email{andrea.simoni@univ-rennes1.fr}
\affiliation{Univ Rennes, CNRS, IPR (Institut de Physique de Rennes)-UMR 6251, F-35000 Rennes, France}
\author{M. Zaccanti}
\affiliation{Istituto Nazionale di Ottica del Consiglio Nazionale delle Ricerche (CNR-INO), 50019 Sesto Fiorentino, Italy}
\affiliation{\mbox{European Laboratory for Non-Linear Spectroscopy (LENS), Universit\`{a} di Firenze, 50019 Sesto Fiorentino, Italy}}


\begin{abstract}
We investigate ultracold collisions in a novel mixture of $^6$Li and $^{53}$Cr fermionic atoms, discovering more than 50 interspecies Feshbach resonances  via loss spectroscopy. Building a full coupled-channel model, we unambiguously characterize the $^{6}$Li-$^{53}$Cr scattering properties and yield predictions for other isotopic pairs.
In particular, we identify various Feshbach resonances that enable the controlled tuning of elastic $s$- and $p$-wave $^{6}$Li-$^{53}$Cr interactions. Our studies thus make lithium-chromium mixtures emerge as optimally-suited platforms for the experimental search of elusive few- and many-body regimes of highly-correlated fermionic matter, and for the realization of a new class of ultracold polar molecules possessing both electric and magnetic dipole moments.
\end{abstract}

\maketitle
Strongly interacting mixtures of two different kinds of fermionic particles \cite{Casabuoni2004,Hammer2013,Wang2018,Jiang2021} exhibit a plethora
of few- and many-body phenomena extending well beyond single-component systems. 
In particular, this holds true for ultracold mixtures of chemically different atomic species, 
for which strong interactions, combined with a \textit{heavy-light} mass imbalance $M/m$, are predicted to promote a variety of exotic regimes hard, or even impossible to attain with their homonuclear counterparts. 
Notable examples are exotic few-particle clusters \cite{Efimov1973,Castin2010,Endo2011,Levinsen2011,Endo2012,Bazak2017,Kartavtsev2007,Nishida2008,Levinsen2009,Ngampruetikorn2013,Levinsen2013,Blume2012,Bazak2017b, Liu2022A} expected to develop for peculiar mass asymmetries $8\!\lesssim\!M/m\!\lesssim\!14$ \cite{Endo2011,Levinsen2011,Endo2012,Bazak2017,Kartavtsev2007}, novel types of impurity physics \cite{Mathy2011,Massignan2014,Schmidt2018,Liu2022B} and mediated quasi-particle interactions \cite{Bulgac2006,Suchet2017,Camacho-Guardian2018}, which promote the emergence of elusive many-body states \cite{Gubbels2009,Gubbels2013,Endo2016,Pini2021}.
Moreover, weakly-bound bosonic dimers created from a Fermi mixture,
thanks to their increased collisional stability \cite{Petrov2004,Jag2016} near an $s$-wave Feshbach resonance (FR) \cite{Chin2010}, represent an unparalleled starting point to realize degenerate samples of ground-state polar molecules \cite{Ni2008,Takekoshi2014,Molony2014,Park2015,Guo2016,Son2020}.
This has stimulated the search for heteronuclear Fermi systems with suitable Feshbach resonances, resulting in pioneering studies of the bi-alkali $^6$Li-$^{40}$K combination \cite{Wille2008,Voigt2009,Costa2010,Naik2011,Trenkwalder2011,Kohstall2012,Jag2014,Cetina2015,Cetina2016}, and by more recent investigation of $^{40}$K-$^{161}$Dy \cite{Ravensbergen2020} and $^{6}$Li-$^{173}$Yb \cite{Green2020} alkali-lanthanide mixtures. 

Here, we explore ultracold collisions in a novel $^6$Li-$^{53}$Cr Fermi mixture of alkali and transition metal atoms, which is uniquely interesting for two main reasons: 
First, the peculiar chromium-lithium mass ratio, $M/m\!\sim\!8.8$, is optimally-suited to unveil novel non-Efimovian few-body states \cite{Kartavtsev2007,Endo2011,Levinsen2011,Endo2012,Bazak2017}. Notably, these are predicted to be collisionally-stable,  
implying that their presence within a many-particle system will not lead to a reduced lifetime \cite{Levinsen2011, Endo2016, Jag2016}. 
This may open the unparalleled possibility to resonantly increase \cite{Levinsen2011,Endo2012,Bazak2017} non-perturbative, \textit{elastic} few-body effects, and to probe their impact at the many-body level.
 Second, recent studies \cite{Zaremba2022} predict LiCr ground-state dimers to combine a large electric dipole moment exceeding 3 Debye with a $5/2$ electronic spin, making Li-Cr mixtures appealing candidates to realize ultracold polar, paramagnetic molecules. 
Yet, whether Li-Cr mixtures exhibit favorable scattering properties, crucial to access these compelling scenarios, has been so far unpredictable due to the complete lack of experimental input.

We positively answer this question by identifying various $s$- and $p$-wave Feshbach resonances well-suited for the controlled tuning of elastic Li-Cr interactions.
Through loss spectroscopy measurements \cite{Chin2010} we reveal more than 50 isolated inter-species FRs, arranged in non-chaotic patterns despite the dipolar nature and complex level structure of fermionic $^{53}$Cr \cite{Neri2020}, 
reminiscent of $^{161}$Dy \cite{Lu2012} and $^{167}$Er \cite{Aikawa2014}. 
We construct a full coupled-channel model \cite{Chin2010} able to unambiguously connect the observed features to well-defined LiCr molecular states. 
%
We experimentally characterize one among such resonances, finding excellent agreement with our model predictions. We also find favorable scenarios for other isotopic pairs, opening the route to future few- and many-body studies with resonantly-interacting lithium-chromium mixtures.

By following procedures reported elsewhere \cite{Neri2020,Ciamei2022}, we produce ultracold samples of $^6$Li and $^{53}$Cr atoms in a
bichromatic optical dipole trap (BODT), formed by a multi-mode infrared (IR) laser beam centered at 1073 nm \cite{Simonelli2019},
overlapped with a green beam at 532 nm, focused to a waist of 45 $\mu$m and 40 $\mu$m, respectively. The lithium-to-chromium trap depth ratio set by
the primary IR source, of about 3, can be reduced by the green light, that confines chromium and anti-confines lithium, allowing to
controllably adjust the Li and Cr populations, densities, and temperature.
To perform FR scans, we employ ultracold mixtures at 10(1) $\mu$K, trapped in the sole IR beam with negligible differential gravitational sag, comprising about $10^5$ Cr and 1.8 $10^6$ Li atoms, characterized by Cr (Li) peak densities of $10^{11}$ cm$^{-3}$ (2 $10^{12}$ cm$^{-3}$).
Spin-state manipulation \cite{Ciamei2022} allows us to explore different binary Li-Cr mixtures: lithium is prepared in either of the
two lowest Zeeman levels of the electronic and hyperfine ground-state manifold $^2S_{1/2}$, $|f_{\rm Li}\!=\!1/2, m_{f,\rm Li}\!=\pm\!1/2\rangle$,
hereafter labeled Li$|1\rangle$ and Li$|2\rangle$, respectively \cite{footnote}. 
Chromium, initially produced in the lowest hyperfine and Zeeman level $|f_{\rm Cr}\!=\!9/2, m_{f,\rm Cr}\!=\!-9/2\rangle$ of its electronic ground state $^7S_{3}$ \cite{Neri2020} (hereafter denoted
Cr$|1\rangle$), can be also transferred to the two higher-lying Zeeman states of the $f_{\rm Cr}\!=\!9/2$ manifold, labeled Cr$|2\rangle$ and Cr$|3\rangle$, respectively.
 We explore all six Li$|i \rangle$-Cr$|j \rangle$ mixtures with $i \!=1,2$ and $j \!=1,2,3$, each being characterized by the total spin projection quantum
number, $M_f\!=m_{f,Li}+m_{f,Cr}\!=-i+j-4$, thus spanning $-5\! \leq M_f\! \leq\!-2$.
\begin{figure}
\begin{center}
\includegraphics[width=\columnwidth]{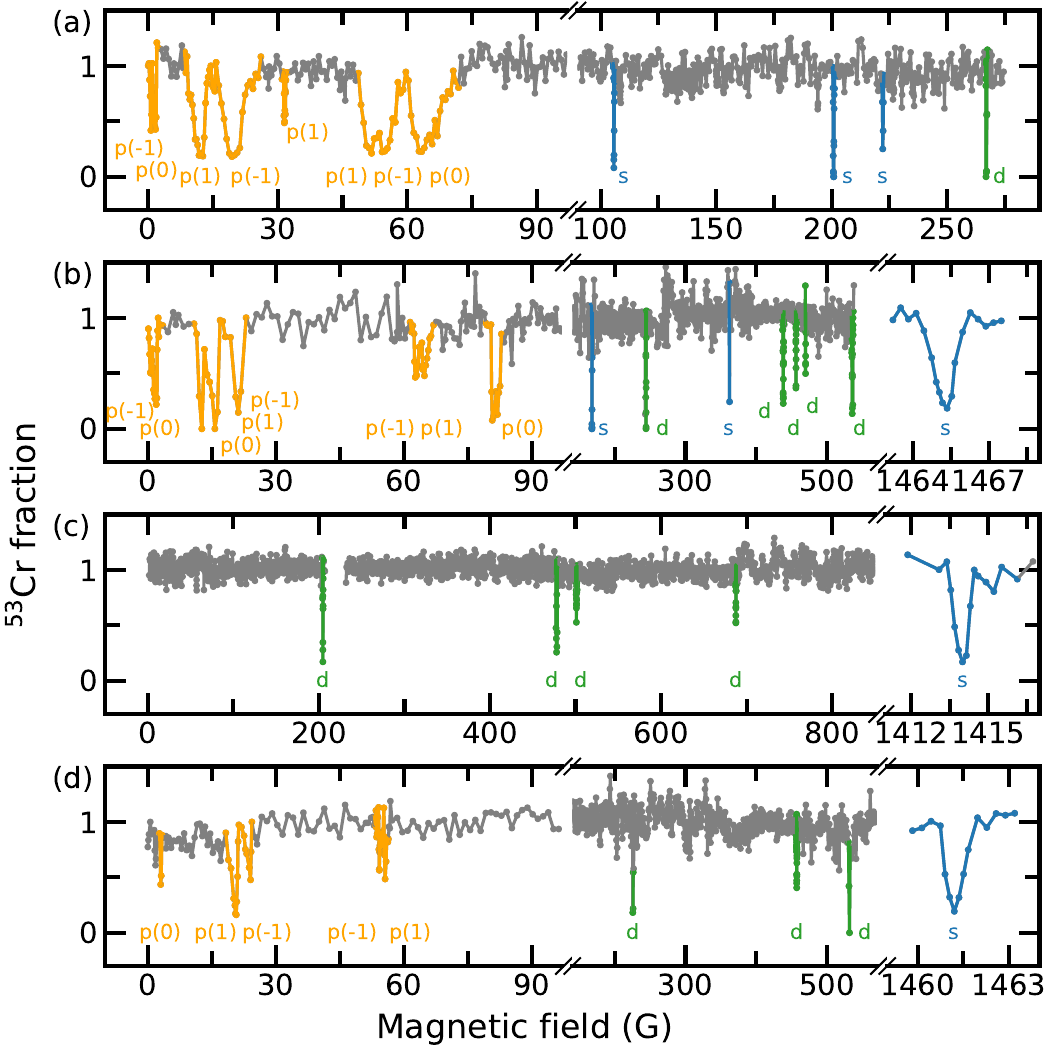}
\vspace*{-20pt}
\caption{Overview of $^{6}$Li-$^{53}$Cr loss spectra. The remaining Cr number, normalized to its background value, recorded after an interaction time $t_H$ with lithium, is plotted as a function of the magnetic field for four different combinations: \textbf{(a)} Li$|2\rangle$-Cr$|3\rangle$, $M_f\!=-3$; \textbf{(b)} Li$|2\rangle$-Cr$|2\rangle$, $M_f\!=-4$; \textbf{(c)} Li$|1\rangle$-Cr$|1\rangle$, $M_f\!=-4$; \textbf{(d)} Li$|2\rangle$-Cr$|1\rangle$, $M_f\!=-5$. Each point is the average of at least four independent measurements. $t_H\!=$4 s for all but the (c) panel, where $t_H\!=$5 s. Features that our model links to $s$-, $p$- and $d$-wave molecular levels are colored blue, orange and green, respectively. Numbers in brackets indicate the assigned $m_l$.
}
\label{Fig1}
\end{center}
\vspace*{-10pt}
\end{figure}

We perform loss spectroscopy through magnetic-field scans with typical step size of 60 mG. We keep the sample at a variable field for a
fixed time $t_H$, and then monitor the remaining atoms via spin-selective absorption imaging \cite{Ciamei2022}. FRs
are identified by enhanced atom losses, induced by different processes depending on the spin combination investigated \cite{Chin2010}.
Any Li$|i\rangle$-Cr$|j\rangle$ mixture but the lowest-energy one, $i$=$j$=1, may undergo two-body losses.  Inelastic spin-exchange occurs
whenever the initial atom pair is coupled to an energetically lower channel with equal $M_f$ and orbital partial wave $l$. Such process
does not affect excited spin combinations with either Li or Cr in the ground state ($i$ or $j$=1).  Weaker dipolar relaxation processes
are enabled by spin-dipole coupling \cite{Chin2010} for any excited mixture, that can decay to lower-lying states with different $M_f$ or
$l$, provided that $\Delta l\!= 0,\pm 2$ and $M_f + m_l$ is conserved, $m_l$ being the projection of $l$ along the magnetic-field
quantization axis \cite{Naik2011}. Three-body recombination processes affect any 
 mixture.
While these can in
principle involve either two light and one heavy atom or vice versa, only the former case is relevant here, given our Li-Cr density imbalance.

Figure~\ref{Fig1} provides an overview of FR scans for four of the six Li-Cr combinations explored herein, from which important
insights can be gained. None of the mixtures exhibits a dense FR spectrum, in contrast with mixtures of alkalis and Er or Dy lanthanides
\cite{Ravensbergen2020,Schafer2021}. Similarly to bi-alkali systems, see e.g. Ref. \cite{Wille2008}, none of the loss peaks is due to, or overlapped with, Cr-Cr or Li-Li resonances. 
Additionally, nearly-coincident FR locations found among different spectra, both for mixtures with equal $M_f$ values (see Fig.~\ref{Fig1}(b) and (c)), and unequal ones, suggest that relatively few molecular states, split into different hyperfine levels, suffice to explain our observation.
 Finally, while few, sparse and narrow features characterize the high-field spectral regions above 150 G, each scan but the Li$|1\rangle$-Cr$|1\rangle$ one exhibits more complex patterns below 150 G, with strong
loss peaks arranged in doublet or triplet structures, see e.g. Fig.~\ref{Fig1}(b). This suggests that three-body losses
are overcome by two-body ones in such field region, and that the observed FRs likely occur in $l\!>0$ partial waves,
split by magnetic dipole-dipole interaction \cite{Ticknor2004,Cui2017} and possibly other couplings \cite{Zhu2019}.

\begin{figure}
\begin{center}
\includegraphics[width=89mm]{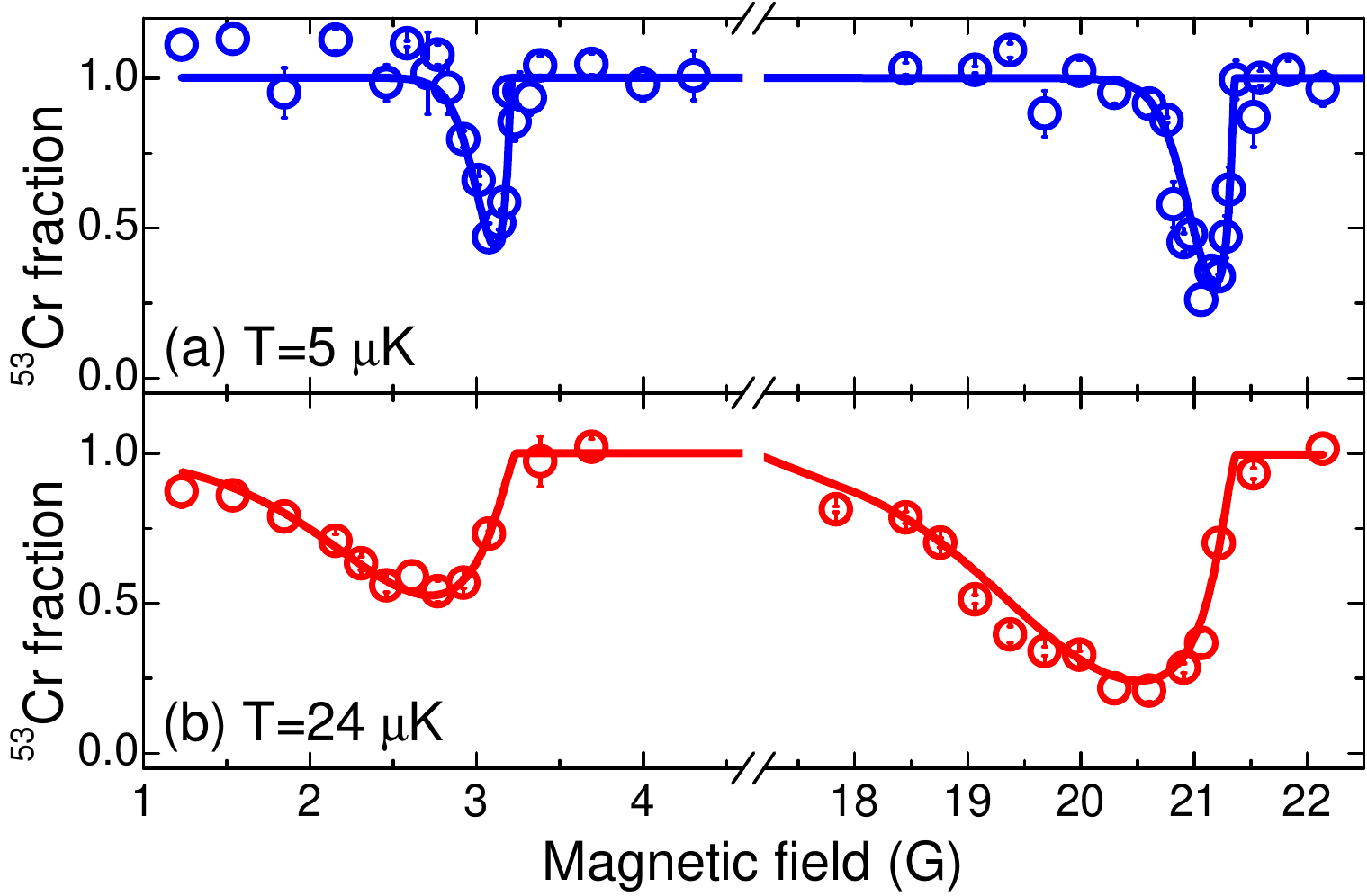}
\vspace*{-20pt}
\caption{
Temperature dependence of atom loss at two low-field FRs of Li$|2\rangle$-Cr$|1\rangle$. \textbf{(a)} The remaining Cr fraction,
recorded after $t_H$=4 s with a Li sample at 5~$\mu$K, is plotted as a function of the magnetic field. \textbf{(b)} Same as panel
(a) but for 24 $\mu$K. Each data point is the average of three independent measurements. Lines are best fits to the
model of Ref.~\cite{Jones1999} for $p$-wave collisions.}
\label{Fig2}
\end{center}
\vspace*{-10pt}
\end{figure}
This intuition is supported by additional loss data, recorded near two low-field Li$|2\rangle$-Cr$|1\rangle$
FRs on low-density samples that were prepared at 5 $\mu$K and 24 $\mu$K, shown in Fig.~\ref{Fig2}(a) and (b), respectively.
Both features are highly asymmetric and they sensitively widen for increasing temperature, a behavior typical of $l\!>\!0$-wave FRs
\cite{Ticknor2004,Cui2017,Fouche2019}. Without aiming at a quantitative lineshape analysis, we remark that our data are
well reproduced (see solid lines in Fig.~\ref{Fig2}) by a simple model \cite{Jones1999} that assumes $l\!=\!1$ collisions and solely accounts
for two-body processes \cite{NoteK2}. 
Another peculiarity of Fig.~\ref{Fig2} spectra, common to other features in Fig.~\ref{Fig1}, is their thermal tail \cite{Ticknor2004,Cui2017,Fouche2019} oriented towards lower fields. This implies that such FRs originate from ``anomalous" molecular levels that lay slightly above the atomic threshold at zero
field and cross it with negative differential magnetic moments.

We now  move to quantitatively analyze our spectroscopic data building a full coupled-channel model \cite{Chin2010}. This accounts for the atomic hyperfine and Zeeman energies \cite{Childs1963,Arimondo1977}, defining the asymptotic collision thresholds. The electrostatic interaction is represented by adiabatic Born-Oppenheimer potentials of \textit{sextet} $X ^6\Sigma^+$ and \textit{octet} $a ^8\Sigma^+$ symmetry, where $S\!=5/2$ and $7/2$ denote the total electron spin of LiCr dimers. 
Such potentials, correlating at long range with the well-known analytical form \cite{Chin2010} comprising a multipolar expansion plus the exchange potential, are  parametrized in terms of sextet $a_6$ and octet $a_8$ $s$-wave scattering lengths, respectively.
Besides this strong isotropic interaction, 
both $l$- and $M_f$-conserving, 
we also account
for weaker anisotropic couplings originating from the long-range magnetic spin and short-range second-order spin-orbit interactions. These  terms can couple different partial waves and hyperfine states, with the selection rules discussed above for dipolar relaxation processes \cite{Chin2010,SM}. 

\begin{table}
\centering
\begin{tabular}{l|c|c|c|c|c|c}
\hline
\hline
 $i,j,l,m_l$  & $B_{exp} $  & $B_0 $ & $S_r,l_r$ & $ a_{\rm bg}^{(l)} $ & $\Delta_{\rm el}$ & $\delta \mu $  \\ 
                      & $ (G)$      & $(G)$  &             & $(a_0^{2l+1})$    &   $ (G)$    & $ (\mu_B)$      \\ 
\hline			
 $1,1,0,0$  & $204.6$  & $204.7$  & $5/2,2$     & $41.3$ & $7.0 \,10^{-3}$   & $3.7$    \\ 
 $1,1,0,0$  & $477.6$  & $478.1$  & $5/2,2$     & $41.5$ & $1.8 \,10^{-3}$   & $2.0$    \\
 $1,1,0,0$  & $501.0$  & $501.9$  & $5/2,2$     & $41.5$ & $3.8\,10^{-4}$     & $2.0$    \\
 $1,1,0,0$  & $687.4$  & $687.1$  & $5/2,0$     & $41.5$ & $2.3 \,10^{-4}$     & $4.0$    \\
 $1,1,0,0$  & $1414.0$ & $1414.1$  & $5/2,0$     & $41.5$ & $0.47$      & $2.0$    \\
\hline
 $2,1,1,0$ & $3.05$  & $2.3$ & $7/2,1$ &  $-1.6\, 10^{5}$ & $-3.70$ & $-0.56$   \\ 
 $2,1,1,1$ & $21.1$  & $20.9$ & $7/2,1$ &  $8.2\, 10^{3}$ & $119$ & $-0.28$   \\ 
 $2,1,1,-1$ & $24.2$  & $24.2$ & $7/2,1$ &  $2.5\, 10^{4}$ & $37.8$ & $-0.24$ \\
 $2,1,1,-1$ & $54.3$ & $54.8$ & $7/2,1$ &  $1.4\, 10^{5}$ & $-4.56$ & $0.27$  \\
 $2,1,1,1$ & $55.6$   & $56.1$ & $7/2,1$ &  $-1.1\, 10^{5}$ &$5.08$ & $0.31$   \\
 $2,1,0,0$ & $225.7$  & $225.8$ & $5/2,2$ &  $41.3$ & $7.4\,10^{-3}$ & $3.8$  \\
 $2,1,0,0$ & $457.0$ & $456.7$ & $5/2,2$ &  $41.5$ & $3.6\,10^{-4}$ & $2.0$  \\ 
 $2,1,0,0$ & $531.4$  & $531.8$ & $5/2,2$ &  $41.5$ & $2.3\,10^{-4}$ & $2.0$  \\
 $2,1,0,0$ & $1461.2$ & $1461.2$ & $5/2,0$ &  $41.5$ & $0.48$ & $2.0$  \\
\hline
 $1,2,0,0$ & $65.0$ & $65.9$ & $7/2,0$ &  $39.5$ & $6.6\,10^{-3}$ & $3.1$   \\ 
 $1,2,0,0$ & $135.7$ & $135.7$ & $5/2,2$ &  $40.8$ & $3.7\,10^{-5}$  & $5.0$    \\
 $1,2,0,0$ & $139.5$ & $140.4$ & $7/2,0$ &  $40.8$ & $1.9\,10^{-3}$ & $3.0$    \\
 $1,2,0,0$ & $483.5$ & $484.2$ & $5/2,2$ &  $41.5$ & $1.7\,10^{-2}$ & $2.0$  \\
 $1,2,0,0$ & $1418.1$ & $1417.9$ & $5/2,0$  & $41.5$ & $0.47$ & $2.0$   \\
\hline
\end{tabular}
\vspace*{-5pt}
\caption{
Selection of $^6$Li-$^{53}$Cr Feshbach resonances immune to inelastic spin-exchange decay. Experimental $B_{exp}$ and theoretical $B_0$ locations
are compared for Li-Cr $(i,j)$ channels. 
$B_{exp}$ are obtained from the zeroes of the
numerically-computed first derivative of loss data. 
A conservative 200 mG uncertainty combines our 10 mG field stability over 4 s with day-by-day field drift and systematic errors.
For incoming $s$-wave ($p$-wave) FRs, background scattering lengths $a_{\rm bg}^{(0)}$ (volumes $a_{\rm bg}^{(1)}$) and magnetic widths $\Delta_{\rm el}$ are computed in the zero-energy limit (at collision energy $E/k_B=10 \mu K$). 
 Coupled $s$- and $d$-waves ($p$-wave only) are included for $s$-wave ($p$-wave) resonances.
The differential magnetic moment $\delta \mu$, the electron spin $S_r$, and the rotational angular momentum  $l_r$ of the molecular state are also listed.
Additional data are reported in \cite{SM}.
} 
\label{Table1}
\vspace*{-5pt}
\end{table}

Starting from the {\it ab-initio} data and long-range potential parameters of Ref.~\cite{Jeung_2010}, we optimize the initially unknown values of $a_6$ and $a_8$, and of the dispersion coefficients $C_6$ and $C_8$, using least-square iterations by comparison with the experimental data.
By tentatively identifying the isolated loss peaks above 1 kG as spin-exchange $s$-wave
FRs associated with $S\!=\!5/2$ rotationless ($l_r\!=\!0$) molecular states, the sextet scattering length is strongly constrained to $a_6\!=\!15.5~a_0$. This also reproduces several other features at lower field, assigned to $l_r\!=\!2$ states of the $S\!=\!5/2$ potential, thus confirming the hypothesis.
A strong constraint on  $a_8$  is instead provided by the peculiar low-field patterns, characterized by irregularly-spaced triplets and several ``anomalous"FRs (see  Fig.~2).
A global least-square fit to the observed FRs yields the best-fit parameters $a_6\!=\!15.46(15)~a_0$, $a_8\!=\!41.48(2)~a_0$, $C_6\!=\!922(6)~a.u.$, and
$C_8\!=9.8(5)\,10^4~a.u.$, where errors represent one standard deviation obtained from the fit covariance matrix. 
We remark that our $a_8$ value agrees well with recent \textit{ab initio} estimates \cite{Zaremba2022}.

Our quantum collisional model accurately reproduces the experimental findings,
 as shown in Table~\ref{Table1} for a subset of 20 FRs. Besides the magnetic field location, we provide the relevant quantum numbers for both the entrance channel and molecular state, together with the associated resonance parameters: background scattering length $a_{\rm bg}^{(0)}$ (or volume $a_{\rm bg}^{(1)}$), magnetic width $\Delta_{\rm el}$, and magnetic moment $\delta \mu$ of the molecular state relative to the atomic threshold.
The low-field  spectral region is entirely dominated by $p$-wave FRs, featuring  $m_l$ splittings much larger than those found in alkali systems \cite{Ticknor2004,Cui2017,Zhu2019}, owing to the increased role of spin-spin dipole coupling in Li-Cr
mixtures and to the coincidentally small relative magnetic moment of the molecular states involved.
In particular, we highlight the presence in Li$|2\rangle$-Cr$|1\rangle$ of a strong $p$-wave FR with $m_l\!=\!-1$, centered around 24 G and essentially immune to two-body losses \cite{SM}, 
well-suited to investigate $p$-wave resonant Fermi mixtures \cite{Gurarie2005}. 

\begin{figure}
\begin{center}
\includegraphics[width=89mm]{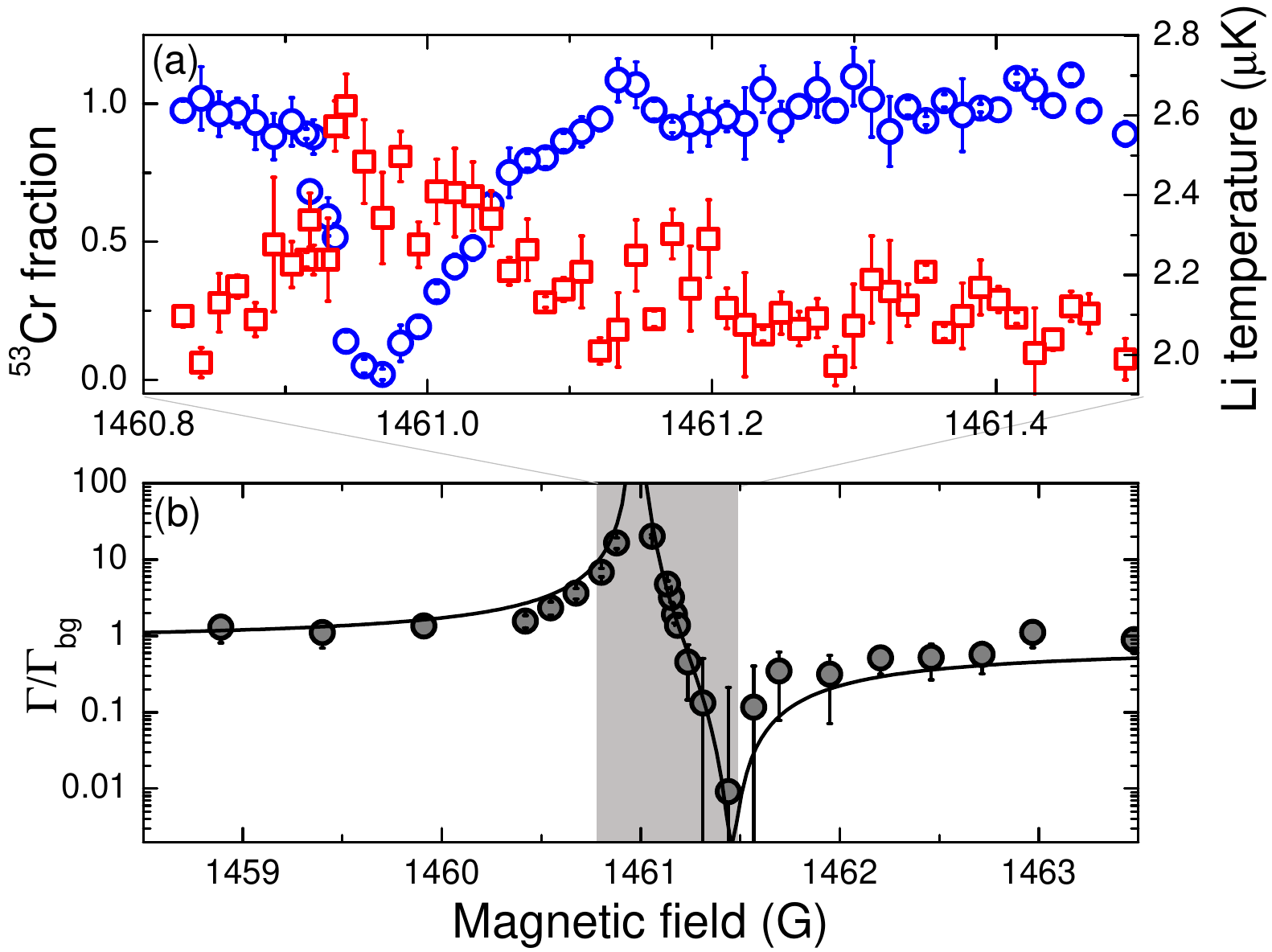}
\vspace*{-20pt}
\caption{
Inelastic and elastic scattering at a Li$|2\rangle$-Cr$|1\rangle$ $s$-wave
FR. \textbf{(a)} Magnetic-field dependence of the remaining Cr fraction (blue circles,
left axis) and final Li temperature (red squares, right axis), recorded
on a mixture at 2 $\mu$K after 150 ms interaction time. Each data point is the average of at
least three independent measurements. The loss peak corresponds to a
$1/e$ lifetime of 40(3) ms. \textbf{(b)} Field dependence of
the collision-induced damping rate $\Gamma$ of a Li cloud at 2 $\mu$K,
sloshing along the weak axis of the BODT in the presence of a Cr cloud at
rest. 
Experimentally determined normalized $\Gamma/\Gamma_{bg}$ is compared with the thermally-averaged elastic scattering rate given by our  model (solid line).  
Error bars account for the statistical error of the fit of the oscillation data to a damped sinusoidal function. The shaded
area marks panel (a) boundaries. For both panels the field uncertainty is 5 mG.}
\label{Fig3}
\end{center}
\vspace*{-10pt}
\end{figure}

We also identify various $s$-wave FRs in different spin-combinations, including the absolute Li-Cr ground state. Owing to
the relatively small values of $a_6$ and $a_8$ \cite{Chin2010}, similarly to the Li-K case \cite{Wille2008,Naik2011}, these features are generally narrow. In particular, our model connects all FRs observed above 1400 G to $l_r\!=\!0$ molecular levels of $X ^6\Sigma^+$ potentials, predicting negligible two-body loss rates, magnetic-field widths $\Delta_{\rm el}\!\sim$0.5~G, and associated effective-range parameters \cite{Chin2010,Kohstall2012} of a few 1000~$a_0$.

We test these  expectations by  inspecting inelastic and elastic properties of the FR located at 1461 G occurring in the
Li$|2\rangle$-Cr$|1\rangle$ channel.
Fig.~\ref{Fig3}(a) shows a 10 mG-resolution loss scan, performed with a mixture initially produced at 2 $\mu$K, comprising about 5 10$^5$ Li and 10$^5$ Cr atoms, respectively, and characterized by a Li (Cr) peak density of about
4 10$^{12}$ cm$^{-3}$ (10$^{12}$ cm$^{-3}$). 
Through a 30 ms-long linear ramp, the magnetic field is lowered from 1464 G to the final value, where the sample is
held for 150 ms. The observed loss feature, shown in Fig.~\ref{Fig3}(a) for the Cr component (blue circles), is strongly asymmetric, as
expected for narrow FRs \cite{Li2018}. Contrarily to the low-field FRs of Fig.~\ref{Fig2}, here we
find a lithium-to-chromium loss ratio consistent with 2 within experimental uncertainty. This points to Li-Li-Cr three-body processes
that overcome two-body ones, in agreement with our model expectation of a small dipolar relaxation rate $K_2\,<10^{-14}$~cm$^3$/s at 2~$\mu$K. Concurrently, we observe a sizable temperature increase of the Li cloud (red squares in Fig.~\ref{Fig3}(a)), pointing to anti-evaporation dynamics due to recombination processes \cite{Chin2010}.

Additionally, we investigate the magnetic-field dependence of the Li-Cr elastic scattering across this FR. By means of experimental protocols
identical to those of Ref.~\cite{Naik2011}, we monitor the damping of the sloshing motion of a lithium cloud of
about 5 10$^4$ atoms, along the weak axis of our BODT, induced by interspecies collisions with about 10$^5$ Cr atoms prepared at 2~$\mu$K \cite{Ciamei2022}.
A species-selective optical excitation \cite{Naik2011} initiates sloshing of the Li component while negligibly affecting the Cr one.
For each field, we trace the Li center-of-mass oscillations, which are fit to a damped sinusoidal function to extract the damping rate
$\Gamma$. For weak interactions, with only few scattering events per oscillation period, $\Gamma$ is directly proportional
to the elastic scattering rate \cite{Naik2011}. 
Fig.~\ref{Fig3}(b) shows the experimentally-determined $\Gamma$, normalized to the background value $\Gamma_{bg}$ measured far from the resonant region (circles), together with the thermally-averaged elastic scattering rate obtained from our collisional model (solid line). 
The remarkable agreement between experimental and theoretical data further demonstrates the accuracy of our theory, confirming the model expectation of $\Delta_{\rm el}\!\sim$0.5 G.
Finally, we exploit our quantum collisional model to inspect the scattering properties of other isotopic pairs. Specifically, we focus on the Fermi-Bose mixtures $^6$Li-$^{52}$Cr and $^7$Li-$^{53}$Cr, obtained by replacing fermionic Cr and Li with their most abundant bosonic isotopes, respectively. Within the
Born-Oppenheimer approximation, this can be achieved by simply changing the reduced mass in the Hamiltonian, although the result depends upon the actual number of vibrational states $N_6$ and $N_8$ of the sextet and octet potentials, which bear an uncertainty of a few units \cite{SM}. Setting them to the nominal values employed in our fit \cite{Jeung_2010}, for the $^6$Li-$^{52}$Cr combination we obtain $a_6^{6-52}\!=19.88(13) a_0$ and $a_8^{6-52}\!=42.83(2) a_0$, finding a few  $\sim$100 mG wide resonances at fields of a few hundred Gauss. Since the reduced mass and, correspondingly, the potential parameters of such isotopic pair are very similar to the $^6$Li-$^{53}$Cr ones, our predictions are only weakly affected by the actual $N_6$ and $N_8$ values \cite{SM}.
The same procedure, applied to $^7$Li-$^{53}$Cr, yields $a_6^{7-53}\!=38.20(14) a_0$ and $a_8^{7-53}\!=51.53(3) a_0$, respectively. Thanks to the complex hyperfine structure of $^{53}$Cr, this isotopic pair exhibits richer FR spectra, and a few-Gauss wide resonances below 500 G, although the increased sensitivity of the mass-scaled parameters of $^7$Li-$^{53}$Cr upon $N_{6,8}$ variations enhances our systematic uncertainty \cite{SM}.  

In conclusion, we thoroughly investigated the collisional properties of a novel $^6$Li-$^{53}$Cr ultracold Fermi mixture.  
All Li-Cr isotopic pairs, characterized by isolated and non-chaotic FRs linked to molecular states with well-defined quantum numbers, 
combine the simplicity of bi-alkali systems with richer molecular structures as those of alkali-lanthanide mixtures \cite{Barbe2018, Ravensbergen2020,Green2020,Schafer2021}. Several strong and isolated features, already identified, with widths exceeding a few 100 mG, provide an optimal starting point to form (bosonic or fermionic) Feshbach dimers. These could be then transferred via coherent optical schemes to the ground states of the $X ^6\Sigma^+$ and $a ^8\Sigma^+$ potentials, both predicted \cite{Zaremba2022} to feature sizable electric and magnetic dipole moments, thus making Li-Cr a competing system to realize ultracold paramagnetic polar molecules. 
The $s$-wave FRs in $^6$Li-$^{53}$Cr (and $^7$Li-$^{53}$Cr), with character similar to Li-K ones \cite{Wille2008,Naik2011} but immune to two-body losses, together with an increased mass ratio, provide an unparalleled opportunity to investigate novel non-Efimovian few-body physics \cite{Kartavtsev2007,Endo2012,Bazak2017,Nishida2008,Levinsen2009}.
The presence of $s$- and $p$-wave FRs, featuring lifetimes of several tens of ms in the resonant region, makes lithium-chromium promising also for future many-body studies.

\begin{acknowledgments}
We thank  D. Petrov, J. Szczepkowski, E. Tiemann, M. Tomza, K. Zaremba-Kopczyk and the LENS Quantum Gases group for fruitful discussions, and G. Modugno for useful suggestions and a critical reading of the manuscript. This work was supported by the ERC through grant no.\:637738 PoLiChroM, by the Italian MIUR through the FARE grant no.~R168HMHFYM, and through the EU H2020 Marie Sk\l{}odowska-Curie grant agreement no. 894442 [fellowship to A.C.].
\end{acknowledgments}

\vspace*{0pt}




\onecolumngrid


\newpage

\begin{center}
\textbf{
Supplemental Material\\[4mm]
\large Exploring ultracold collisions in $^6$Li-$^{53}$Cr Fermi mixtures: Feshbach resonances and scattering properties of a novel alkali-transition metal system}\\

\vspace{4mm}
A.~Ciamei,$^{1,2,*}$ 
S.~Finelli,$^{1,2}$
A.~Trenkwalder,$^{1,2}$
M.~Inguscio,$^{1,2,3}$
A.~Simoni,$^{4,\dag}$
and M.~Zaccanti,$^{1,2}$\\
\vspace{2mm}
{\em \small
$^1$Istituto Nazionale di Ottica del Consiglio Nazionale delle Ricerche (INO-CNR), 50019 Sesto Fiorentino, Italy\\
$^2$\mbox{LENS and Dipartimento di Fisica e Astronomia, Universit\`{a} di Firenze, 50019 Sesto Fiorentino, Italy}\\
$^3$Department of Engineering, Campus Bio-Medico University of Rome, 00128 Rome, Italy\\
$^4$Univ Rennes, CNRS, IPR (Institut de Physique de Rennes)-UMR 6251, F-35000 Rennes, France\\}
{\small$^*$ E-mail: ciamei@lens.unifi.it}\\
{\small$^{\dag}$ E-mail: andrea.simoni@univ-rennes1.fr}

\end{center}

\setcounter{equation}{0}
\setcounter{figure}{0}
\setcounter{table}{0}
\setcounter{section}{0}
\setcounter{page}{1}
\makeatletter
\renewcommand{\theequation}{S.\arabic{equation}}
\renewcommand{\thefigure}{S\arabic{figure}}
\renewcommand{\thetable}{S\arabic{table}}
\renewcommand{\thesection}{S.\arabic{section}}

\section{Extraction of resonance parameters in our collisional model} 

In order to summarize information from numerical calculations,
resonances can be parametrized in terms of few characteristic quantities.
Our starting point is the expression of the generalized energy-dependent
scattering length for $l$-wave collisions $a^{(l)}$. It can be shown
that the latter quantity in the presence of a series of overlapping
inelastic resonances, can always be expressed in the intuitive form
\cite{Simoni2021}

\begin{equation} 
a^{(l)}=a_{\rm bg}^{(l)}+ \sum_k \frac{p_k}{B-B_{0,k}-i \gamma_{B,k}/2} ,
\label{eq_basic}
\end{equation}
with $\gamma_{B,k}$ the inelastic magnetic width and $p_k$ the pole
strength of resonance $k$. The latter can in turn be used in order
to define the elastic magnetic width $\Delta_{{\rm el},k}$ according
to $p_k = - a_{{\rm bg},k}^{(l)} \Delta_{{\rm el},k}$, where $a_{{\rm
bg},k}^{(l)}= a_{{\rm bg}}^{(l)}(B_{0,k})$. In the presence of direct
({\it i.e.} non-resonant) inelastic scattering, the background scattering
length, the pole strength and the elastic magnetic width are all complex
quantities. Reference~\cite{Simoni2021} outlines a sophisticate numerical
procedure to extract such lineshape parameters.

Owing to the inverted hyperfine structure of $^{53}$Cr, all hyperfine
combinations considered in Table~1 of the main text are stable under
spin-exchange collisions and only decay by weaker anisotropic spin
interactions. In this weak-inelasticity scenario, one can neglect the
imaginary parts of $a_{{\rm bg}}^{(l)}$ and $p_k$, and the theory of Ref.~\cite{Simoni2021} reduces to the one of Ref.~\cite{Naik2011}. Following the
formalism described in the latter reference, we characterize isolated resonances
in $s$-wave collisions in terms of two experimentally relevant {\it real}
parameters, the elastic $\Delta_{\rm el}$ and the inelastic $\gamma_B$
magnetic width, evaluated in the zero-energy limit.
The quantity $\gamma_B$ concurs to the magnetic-field dependence of
the complex-valued $s$-wave scattering length $a=\alpha-i \beta$ (we
have put by notational simplicity $a=a^{(0)}$), with real and imaginary parts given
respectively by :

\begin{eqnarray}
\alpha&=& a_{bg}-a_{res}\frac{\gamma_B (B-B_0)}{(B-B_0)^2+(\gamma_B/2)^2}\:,\label{re_a}\\
\beta&=&2a_{res}\frac{(\gamma_B/2)^2}{(B-B_0)^2+(\gamma_B/2)^2}\:,
\end{eqnarray}
where $a_{res}=a_{bg} \Delta_{\rm el}/\gamma_B$.
Correspondingly, the elastic and inelastic scattering cross sections can be written exactly as \cite{Hutson2007}:
\begin{eqnarray}
\sigma_{el}&=& 4 \pi |a|^2\frac{1}{1+k^2|a|^2+2k\beta}\:,\\
\label{sigmael}
\sigma_{in}&=&\frac{4 \pi \beta}{k}\frac{1}{1+k^2|a|^2+2k\beta}\:.
\label{sigmain}
\end{eqnarray}
From $\sigma_{in}$, one can also determine the inelastic rate coefficient
$k_2 = v \sigma_{in}$ for two particles with relative velocity $v=\hbar
k / m_{\rm red}$. Note from Eq.~(\ref{sigmain}) that the maximum of the
inelastic collision rate $k_{2,{\rm max}}$ evaluated at the resonance
pole for zero collision energy is related to the magnetic width via
$\gamma_B=4 h a_{\rm bg} \Delta_{\rm el} / ( m_{\rm red} k_{2,{\rm
max}})$, with $m_{\rm red}$ the Li-Cr reduced mass.


From Eq.~(\ref{re_a}), one can easily verify that $\alpha$ reaches its
maximum and minimum values, $a_{bg}(1\pm \Delta_{\rm el}/\gamma_B)\sim
\pm a_{res}$, respectively, at magnetic field detunings of $\mp
\gamma_B/2$. As such, for $\gamma_B \ll \Delta_{\rm el}$, the elastic
scattering can be widely tuned across the resonance. Under these
conditions, the elastic width $\Delta_{\rm el}$ represents the
difference between the magnetic-field locations $(\Delta_{\rm el} \pm
\sqrt{\Delta_{\rm el}^2 - \gamma_B^2})/2 $ where $\alpha$ vanishes. To
very good approximation one zero is close to $B_0$, the other is close to
$\Delta_{\rm el}$. Also, in spite of the fact that the inelastic rate on
resonance becomes very large as $k_{2,{\rm max}} \propto \gamma_B^{-1}$,
a magnetic field region exists where one can control the elastic part
of the scattering length without introducing large losses.


\begin{table}[ht!]
\centering
\begin{tabular}{l|c|c}
\hline
\hline
 $i,j,l,m_{l}$  &  $B_0 $ & $\gamma_B$  \\
                      & $ (G)$ & $ (\mu G) $ \\
\hline
 $1,1,0,0$  & $204.7$  & 0 \\
 $1,1,0,0$  & $478.1$  & 0 \\
 $1,1,0,0$  & $501.9$  & 0 \\
 $1,1,0,0$  & $687.1$  & 0 \\
 $1,1,0,0$  & $1414.1$ &0  \\
\hline
 $2,1,1,0$ & $2.3$  & $-3.6 \times 10^2$  \\
 $2,1,1,1$ & $20.9$  & $-1.65 \times 10^3$  \\
 $2,1,1,-1$ & $24.2$  &  $\sim 0$\\
 $2,1,1,-1$ & $54.8$ & $\sim 0$ \\
 $2,1,1,1$ & $56.1$   & $1.5\times 10^2$  \\
 $2,1,0,0$ & $225.8$  & $39$ \\
 $2,1,0,0$ & $456.7$ & $52$ \\
 $2,1,0,0$ & $531.8$  & $8.9 \times 10^4$ \\
 $2,1,0,0$ & $1461.2$ & $4.3 \times 10^{-2}$ \\
\hline
 $1,2,0,0$ & $65.9$  & $5.2$ \\
 $1,2,0,0$ & $135.7$ &  $24 $ \\
 $1,2,0,0$ & $140.4$ & $67 $ \\
 $1,2,0,0$ & $484.2$ & $6.8 \times 10^{2}$ \\
 $1,2,0,0$ & $1417.9$ & $80$ \\ 
\hline
\end{tabular}
\vspace*{5pt}
\caption{
Magnetic width $\gamma_B$ for the Feshbach resonances in Table 1 of the main text with incoming channel $(i,j)$
and partial wave $(l,m_l)$. The field $B_0$ denotes the theoretical resonance position. 
Only $p$-wave collisions are included for incoming angular momentum $l=1$, coupled $s$ and $d$-waves for $l=0$.
}
\vspace*{5pt}
\label{TabSM1}
\end{table}

The extraction of the resonance parameters for $p$-wave collisions is
more involved, since the background contribution to scattering in our
system turns out to vary rapidly with $B$ and a few Li-Cr resonances
are found to be overlapping. We thus adopt the formalism developed in
Ref.~\cite{Simoni2021}, that allows one to extract resonances parameters
without the need of fitting a lineshape to the numerical data. 

An important, often overlooked, point is that in the presence of
non-negligible long-range spin-spin interactions the scattering
volume $a^{(1)}$ does not exist in the zero-energy limit
\cite{Sadeghpour2000}. Even not taking the $E\to 0$ limit, we find
that at typical ultracold temperatures $a_{\rm bg}^{(1)}$ strongly
increases with decreasing energy. Since the pole strength $p_k$ only
weakly depends on energy, the magnetic width will tend to zero in the
zero-energy limit {\it i.e} at zero energy the scattering volume is not
magnetically tunable. Different ways exist in order to obtain a finite
width in this pathological situation. In the present paper we only work
with the energy-dependent generalized scattering length $a^{(1)}(B,E)$,
evaluated at typical experimental temperatures, which is a finite quantity
amenable of the pole decomposition Eq.~(\ref{eq_basic}). As alternative
methods in the spirit of Ref.~\cite{Cui2017}, one could either subtract
the singular zero-energy contribution from the total scattering phase
or set artificially to zero the long-range part of the interaction in
the entrance channel.

Which procedure is to be preferred depends on the information needed
from two-body theory. The approach followed herein has the advantage
of directly providing the energy-dependent scattering length needed,
for instance, in multipolar isotropic or anisotropic pseudopotentials
\cite{Derevianko2003}.  If, on the converse, one chooses to cancel in one
way or the other the effect of the spin-spin interaction, the resulting
scattering amplitude might be used in many-body models as a tunable
short-range (or contact) interaction. However, in this latter case the
spin-spin coupling must then be taken into account at the many-body level,
as customarily done, for instance, in the theory of dipolar Bose-Einstein
condensates.

Table~\ref{TabSM1} provides the inelastic parameter $\gamma_B$, extracted
from our model following the procedure discussed above, for the subset of
selected FRs listed in Table~1 of the main text.  Note that the width $\gamma_B$
does not need to be a positive quantity, the only constraint imposed
by unitarity of the scattering matrix being a non-negative imaginary
part of $a^{(l)}$. As expected, the features occurring in the absolute
ground state of the mixture, Li$|1\rangle$-Cr$|1\rangle$, are immune to
two-body decay processes. Additionally, we remark that also $p$-wave
resonances with $m_l=-1$ occurring in the Li$|2\rangle$-Cr$|1\rangle$
combination exhibit a vanishing $\gamma_B$. This owes to the fact that
such a combination features $M_f=-5$, and thus for these resonances
$M_f+m_l=-6$. Given that the only channel available at lower energy has
$M_f=-4$, and that here only $p$-wave collisions are included, such FRs
have no available inelastic channels conserving $M_F\!=\! M_f+m_l$. Indeed, if
$l=3$ partial waves, i.e. $f$-wave, are accounted for in the calculation
finite dipolar relaxation rate are enabled. Even in this case, however,
the associated two-body loss rate remains at the 10$^{-15}$ cm$^3/s$
level at ultralow temperatures. Such slow atom-loss rate, along with
large magnetic-field widths (see Table ~1 in the main text), makes these
features very appealing for the investigation of few- and many-body
phenomena with $p$-wave Feshbach resonant Fermi mixtures.

Finally, note that the $\gamma_B/\Delta_{\rm el}$ ratio strongly
drops for FR features occurring at higher fields in both
Li$|1\rangle$-Cr$|2\rangle$ and Li$|2\rangle$-Cr$|1\rangle$ spin
combinations. For instance, for the $s$-wave resonance occurring in
Li$|2\rangle$-Cr$|1\rangle$ collisions at 1461 G, and investigated in
detail in Fig.~3 of our paper, exhibits $\gamma_B=4.3 \times 10^{-2}$
$\mu$G, and correspondingly a finite but negligibly small two-body loss rate
coefficient. As such, the associated $a_{res}$ parameter exceeds a few
10$^8\!a_0$, a value about two orders of magnitude higher than the Li-K
FR investigated in Ref.~\cite{Naik2011}, for which $\gamma_B=14$ $\mu$G.


\section{Origin of low-field resonance patterns}

\begin{figure}[t!]
\begin{center}
\includegraphics[width=0.8\columnwidth] {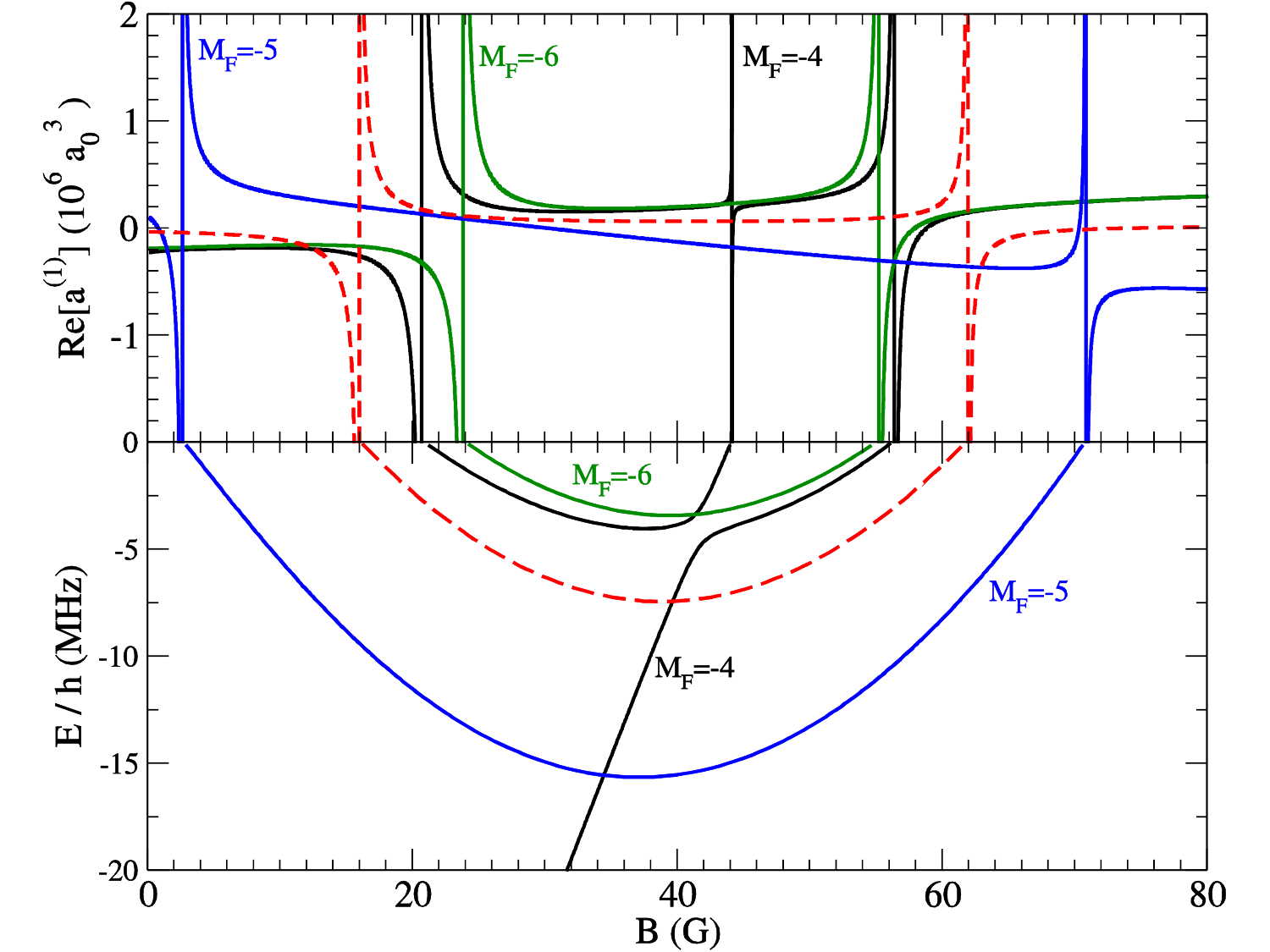}
\caption{
Upper panel : the real part of the energy-dependent $p$-wave scattering volume for
Li$|2\rangle$-Cr$|1\rangle$ collisions for the triplet components
$M_F=-6,-5,-4$. Calculation is performed at a finite collision energy $E/k_B=5\mu$K. 
Two mirror-image triplet patterns are visible with an
extra resonance near 45~G arising from an additional perturbing state.
Lower panel : energies of the corresponding resonant molecular states
with quantum number $M_F$ below the Li$|2\rangle$-Cr$|1\rangle$ atomic
threshold, fixed as zero energy. Pairs of resonances are induced for
each $M_F$ by the {\it same} molecular level crossing twice the atomic
threshold first with negative, then with positive relative magnetic
moment.  In the absence of anisotropic spin dipolar and second-order
spin-orbit interactions the three components would be degenerate and
reduce to the dashed lines in both panels.
} 
\label{fig1_sm}
\end{center}
\end{figure}

The  FR spectra at low magnetic fields (below 100~G), observed in
the Li$|2\rangle$-Cr$|1\rangle$, Li$|2\rangle$-Cr$|2\rangle$ and
Li$|2\rangle$-Cr$|3\rangle$ hyperfine combinations (see Fig.~1 in the
main text), present a peculiar pattern comprising a first low-field
$p$-wave triplet of ``anomalous" resonances (bound state crossing the
atomic threshold from above with increasing $B$) below 30~G, accompanied
by a second ``standard" (bound state crossing the atomic threshold from
below with increasing $B$) resonance triplet, located at higher fields. On
top of these six features, the low-field pattern may include additional
peaks, depending on the specific collision channel.

The origin of the main six-resonance pattern is illustrated in
Fig.~\ref{fig1_sm} with reference to the Li$|2\rangle$-Cr$|1\rangle$
combination. For $p$-wave collisions $l=1, m_l=0,\pm 1$ and the three
projections $-4,-5,-6$ of the exactly conserved quantum number $M_F$
must be accounted for. Let us first turn artificially to zero all
relativistic interactions, such that the potential is fully isotropic.
With reference to the lower panel (dashed line), one can observe a unique
molecular level crossing the atomic threshold first with negative, then
with positive relative magnetic moment, with relatively small values $\sim\! \pm 0.5 \mu_B$.  The upper panel (dashed line) shows the real
part of the energy-dependent scattering volume, that near the location
where the bound state crosses threshold is tuned to large positive and
negative values.

In the presence of anisotropic interactions, degeneracy is lost and
the single molecular level splits into the three components, shown as solid lines in
Fig.~\ref{fig1_sm}, in turn giving rise to a resonance pattern composed by
a pair of triplet peaks, symmetrically occurring, relative to the center
of the pattern.  Note that in general at low fields the splitting is
non perturbative and one does not expect the two resonances for $m_l=\pm
1$ collisions to occur at the same location, as already documented for
alkalis \cite{Viel2016}, {\it albeit} in a less spectacular way. Given its
high sensitivity to small perturbations, this fragile pattern represents an ideal testing ground
for {\it ab-initio} quantum chemistry calculations.

\section{Form of the isotropic and anisotropic potentials}
We first recall that the isotropic interaction in our model is built by smoothly matching the \textit{ab-initio} $^{6} \Sigma^+$ and $^{8}
\Sigma^+$ molecular potentials of Ref.~\cite{Jeung_2010} to the typical long-range analytical form $V_{\rm LR}(R)$ \cite{Chin2010}, comprising the dispersion and the exchange interactions and given by:
\begin{equation}
\label{eq_lr}
V_{\rm LR}(R)=-\dfrac{C_6}{R^6}-\dfrac{C_8}{R^8} \mp A R^\nu e^{-\beta R}.
\end{equation}
As for the singlet and triplet potentials of bi-alkalis, the minus (plus) sign, with $A\!>$0, refers to the deeper sextet (shallower octet) Li-Cr potential. The amplitude $A$, and the constants $\nu$ and $\beta$, solely depend on the properties of separated atoms.
The fit only weakly constrains the exchange-potential parameters, that are thus kept fixed to the nominal values given in Ref.~\cite{Jeung_2010}.

The connection between the short-range \textit{ab-initio} and the long-range form Eq.~\eqref{eq_lr} is performed near the largest distance $14\sim\!a_0$ at which the Li-Cr quantum chemistry data are available \cite{Jeung_2010}. Details of the matching procedure are largely immaterial. In fact, for an isotropic
model, the scattering properties are fully determined once the $s$-wave scattering lengths or, equivalently, the quantum defects of the adiabatic
potentials $^{6,8} \Sigma^+$ are known, along with the leading dispersion coefficient $C_6$ and, to a lesser extent, $C_8$ \cite{Gao2005}.

In addition to the isotropic potentials, we find that accounting for anisotropic interactions is essential to quantitatively interpret the observed FR spectra. The magnetic dipolar interaction is included in its standard asymptotic form \cite{Mies1996}.
The second-order spin-orbit (SSO) interaction is a correction to the
intermolecular potential arising from indirect spin-orbit coupling through
energetically distant electronic states \cite{Mies1996}.  This coupling
only occurs when the electronic clouds of the atoms significantly overlap,
and decreases rapidly with the interatomic distance.  To our knowledge
no information from quantum chemistry is known to date about the SSO
interaction in Li-Cr. For heavy alkalis, the radial dependence of the
SSO has been modeled as a biexponential, the sum of two exponential
functions of different range and amplitude \cite{Kotochigova2000}. In
the absence of any information, we choose here to use a simpler single
exponential form as in early studies \cite{Mies1996}. 
In general, it can be shown that for $\Sigma$ states the second-order
spin-orbit Hamiltonian presents the same spin structure as the usual
spin-spin interaction. Our model SSO for two atoms separated by a vector
distance $\bf R$ reads therefore :
\begin{equation}
V_{SO}=-C \alpha^2 e^{-B(R-R_S)} \left[ \bf{s}_{\rm Li} \cdot \bf{s}_{\rm Cr} - 3 ( \bf{s}_{\rm Li} \cdot {\hat{\bf R}} ) 
( \bf{s}_{\rm Cr} \cdot \hat{\bf R} ) \right], 
\label{SSO}
\end{equation}
where $\alpha$ is the fine structure constant. We fix range parameters
$B=0.83~ a_0^{-1}$ and $R_S=10~ a_0$ of the same order as for alkali
dimers \cite{Mies1996,Kotochigova2000}. For the purposes of the present
work, we do not expect their precise values to be of critical importance.
Eq.~(\ref{SSO}) can also describe a varying direct spin-spin coupling constant
due to electron screening.
The overall amplitude is adjusted to give the best theory-experiment
agreement, with best-fit value $C=1.41\times 10^{-4}$ a.u..  Note that the
SSO contribution is of opposite sign as compared to the magnetic spin
interaction. The SSO is mostly important to model the low-field $p$-wave
multiplet splittings, where it provides about 15\% of the total correction
to the binding energy due to anisotropic interactions. Given the small
value of the relative magnetic moments, this can translate into a large
magnetic-field shift.

\section{Additional data for theory-experiment comparison}

\begin{table}[t!]
\centering
\begin{tabular}{l|c|c|c|c}
\hline
\hline
 $i,j,l,m_l;M_{F}$  & $B_{exp} $  & $B_0 $ & $S_r,l_r$ & $\delta \mu $  \\
                      & $ (G)$      & $(G)$  &              & $ (\mu_B)$      \\
\hline
 $2,2,1,-1;-5$  & $0.73$  &  $0.64$    & $7/2,1$       &      \\
 $2,2,1,0;-4$   & $1.95$  &  $1.81$    & $7/2,1$       &      \\
 $2,2,1,0;-4$  & $12.87$  &  $13.05$   & $7/2,1$       &      \\
 $2,2,1,1;-3$  & $15.45$  &  $16.09$   & $7/2,1$       &      \\
 $2,2,1,-1;-5$  & $21.10$  &  $22.41$   & $7/2,1$       &      \\
 $2,2,1,-1;-5$  & $62.80$  &  $62.64$   & $7/2,1$       &      \\
 $2,2,1,1;-3$  & $64.85$  &  $62.61$   & $7/2,1$       &      \\
 $2,2,1,0;-4$  & $80.94$  &  $81.22$   & $7/2,1$       &      \\
 $2,2,0,0;-4$  & $167.80$ &  $168.58$  & $7/2,0$       & $3.1$     \\
 $2,2,0,0;-4$  & $244.19$  & $244.24$  & $5/2,2$       & $3.7$     \\
 $2,2,0,0;-4$  & $361.93$  & $363.15$  & $7/2,0$       & $1.9$     \\
 $2,2,0,0;-4$  & $437.97$  & $438.98$  & $5/2,2$       & $1.9$     \\
 $2,2,0,0;-4$  & $455.93$  & $455.45$  & $5/2,2$       & $2.0$     \\
 $2,2,0,0;-4$  & $469.5$  &  $470.32$  & $5/2,2$       & $2.0$     \\
 $2,2,0,0;-4$  & $535.37$  & $535.91$  & $5/2,2$       & $2.0$     \\
 $2,2,0,0;-4$  & $1465.30$ & $1464.70$ & $5/2,0$       & $2.0$  \\
\hline
 $2,3,1,-1;-4$  & $0.61$  & $0.64$    & $7/2,1$       &      \\
 $2,3,1,0;-3$  & $1.56$  &  $1.54$    & $7/2,1$       &      \\
 $2,3,1,1;-2$  & $12.17$  & $12.31$    & $7/2,1$       &      \\
 $2,3,1,-1;-4$  & $19.81$  & $20.38$    & $7/2,1$       &      \\
 $2,3,1,1;-2$   &    &   $21.62$   & $7/2,1$       &      \\
               & $31.42$  &     & $7/2,1$       &      \\
 $2,3,1,1;-2$  & $31.57$  & $31.53$    & $7/2,1$       &      \\
               & $31.79$  &     & $7/2,1$       &      \\
 $2,3,1,1;-2$  & $51.47$  & $50.31$     & $7/2,1$       &      \\
 $2,3,1,-1;-4$ & $54.84$  & $54.24$    & $7/2,1$       &      \\
 $2,3,1,0;-3$  & $63.37$  & $62.58$    & $7/2,1$       &      \\
 $2,3,0,0;-3$  & $105.75$  & $106.38$    & $7/2,0$       &   $3.7$   \\
 $2,3,0,0;-3$  & $200.88$  & $201.53$    & $7/2,0$       &   $3.2$   \\
 $2,3,0,0;-3$  & $222.20$  & $222.81$    & $7/2,0$       &   $3.3$   \\
 $2,3,0,0;-3$  & $266.99$  & $266.99$    & $5/2,2$       &   $3.7$   \\
\hline
 $1,3,0,0;-2$  & $194.14$  & $194.82$  & $7/2,0$  &  $3.2$    \\
\hline
\end{tabular}
\vspace*{-5pt}
\caption{
Survey of the Feshbach resonances used to optimize our model, in addition to the ones presented in the main text.
} 
\label{Table1_sm}
\end{table}

We reproduce in Table~\ref{Table1_sm} the additional Feshbach resonance
data we employed to optimize the collision model. Note that channels
where neither particle is in the ground state are not immune under
spin-exchange collisions. Moreover, the number of outgoing inelastic
channels tends to increase as $|M_f|$ decreases. As compared to collision
channels in Table~1 of the main text, the scattering features listed here
are therefore more lossy. We limit ourselves to providing a comparison
of the theoretical and experimental positions, the quantum numbers
of the resonant molecular state, and the relative
magnetic moment for rotationless $l_r\!=0$ molecules. Agreement is
generally very satisfactory, with an average statistical error of $\sim
0.5$~G. Note two $p$-wave resonances predicted by the model near 20~G in
Li$|2\rangle$-Cr$|3\rangle$ collisions are not resolved experimentally
due to thermal broadening. The origin of the resonance triplet at 31~G
in the same spin combination is not yet explained.

We remark that $S_r$ indeed represents a well-defined quantum number for our system at all magnetic fields. This arises from the fact that, since $a_6$ and $a_8$ are sufficiently different, last-below-threshold octet and sextet vibrational levels are also sufficiently far in energy, such that hyperfine mixing is small.
At large magnetic fields, where the Zeeman interaction becomes dominant, also the projection $M_{S_r}$ of the electronic spin on the quantization axis tends to become a good quantum number. While $M_{S_r}$ is not explicitly given, its value can be deduced, if needed, from the differential magnetic moments $\delta \mu$ listed in Table 1 of the main text and in Table \ref{Table1_sm}.
Conversely, owing to the competition of the hyperfine term with the Zeeman interaction and to the presence of the anisotropic dipolar potential, the total
hyperfine spin, \textit{i.e.} the sum of all electron and nuclear spins, is in general not conserved. 

\section{Scattering lengths of $^6$Li-$^{52}$Cr and $^7$Li-$^{53}$Cr isotopic pairs}

We provide here the $s$-wave scattering lengths $a_{6,8}$ for the
adiabatic Born Oppenheimer potentials $X ^6\Sigma^+$ and $a ^8\Sigma^+$
of two isotopic pairs, $^6$Li-$^{52}$Cr and $^7$Li-$^{53}$Cr, of potential
experimental interest.  Since the number of vibrational levels is not
precisely known, we present in Table~\ref{Table2_sm} our results as a
function of the variation $\delta N_{6,8}$ starting with the number of
bound states $N_6=39$ and $N_8=13$ supported by our nominal potentials. To
encompass recent {\it ab-initio} calculations \cite{Zaremba2022} that
predict significantly deeper potentials as our nominal ones, we assume
maximum variations $|\delta N_6|=5$ and $|\delta N_8|=3$.
As already mentioned in the main text, the largest uncertainty due to variations in $\delta N$ concerns
the $^7$Li-$^{53}$Cr system. This can be expected since this isotopic
combination presents the largest variation of reduced masses with respect
to the reference pair $^6$Li-$^{53}$Cr, and thus in the corresponding
scattering phase when one bound state is removed or added from the
potential \cite{Gribakin1993}. On the converse, the reduced mass of
$^6$Li-$^{52}$Cr is close to the one of $^7$Li-$^{53}$Cr, the mass-scaled
potentials for the two isotopes are similar, and our predictions bear
a small dependence upon $\delta N$ variations.

\begin{table}[ht!]
\centering
\begin{tabular}{c|c|c|c|c}
\hline
\hline
 $\delta N_{6,8}$  & $a_6^{6-52} (a_0)$  & $a_8^{6-52} (a_0)$ & $a_6^{7-53}(a_0)$ & $a_8^{7-53}(a_0)$  \\
\hline
    -5              &     $19.34(14)$   &      $$      &     $150(2)$    &    $$   \\
		-4              &     $19.45(14)$   &      $$      &     $92.9(6)$    &    $$   \\
		-3              &     $19.56(14)$   &      $42.47(2)$      &     $69.4(3)$    &    $151.3(3)$   \\
		-2              &     $19.67(14)$   &      $42.62(2)$      &     $55.9(2)$    &    $81.03(7)$   \\
		-1              &     $19.77(14)$   &      $42.73(2)$      &     $46.1(2)$    &    $62.98(4)$   \\
		 0              &     $19.88(13)$   &      $42.83(2)$      &     $38.20(14)$    &    $51.53(3)$   \\
		1              &     $19.99(13)$   &      $42.94(2)$      &     $30.87(13)$    &    $42.86(2)$   \\
		2              &     $20.09(13)$   &      $43.04(2)$      &     $23.49(14)$    &    $35.41(2)$   \\
    3              &     $20.20(13)$   &      $43.15(2)$      &     $15.1(2)$    &    $28.31(2)$   \\
    4              &     $20.31(13)$   &      $$      &     $4.6(2)$    &    $$   \\
    5              &     $20.41(13)$   &      $$      &     $-11.0(4)$    &    $$   \\
\hline
\end{tabular}
\vspace*{-5pt}
\caption{
Sextet and octet $s$-wave scattering lengths $a_{6,8}$ for the
isotopic pairs $^6$Li-$^{52}$Cr and $^7$Li-$^{53}$Cr, predicted from
our $^6$Li-$^{53}$Cr model by mass-scaling. Data are provided for a variation
$\delta N_{6,8}$ in the number of vibrational bound levels supported
by the $^{6,8}\Sigma^+$ adiabatic potentials, counted from the nominal
values $N_6=39$ and $N_8=13$ of Ref.~\cite{Jeung_2010}.
} 
\label{Table2_sm}
\vspace*{-5pt}
\end{table}


\begin{thebibliography}{70}
\bibitem [{foo()}]{footnote}%
  \BibitemOpen
  \href@noop {} {}\bibinfo {note} {With $m_{f,a}$ we denote the projection
  quantum number for element $a$=Li, Cr.}\BibitemShut {Stop}%
\bibitem [{Not()}]{NoteK2}%
  \BibitemOpen
  \href@noop {} {}\bibinfo {note} {Each data set is fitted assuming exponential
  decay of the Cr fraction with loss rate $\gamma(B)\!=\!n_{Li}K_2\!=\!A
  (B_0-B)^{3/2} \mathrm{Exp}[-\frac{\delta \mu (B_0-B)}{k_B T}]$. Temperature
  $T$ is fixed to the experimental value. Amplitude $A$, FR center $B_0$, and
  magnetic moment $\delta \mu$ are fitting parameters.}\BibitemShut {Stop}%
\end{thebibliography}
\end{document}